# An asymmetric non-uniform 3-dB directional coupler with 300-nm bandwidth and a small footprint

**HAMED NIKBAKHT[1], MOHAMMAD TALEBI KHOSHMEHR[1], BOB VAN SOMEREN[2], DIETER TEICHRIB[3,4], MANFRED HAMMER[3], JENS FÖRSTNER[3], AND B. IMRAN AKCA[1,*]**

[1]*LaserLaB, Department of Physics and Astronomy, VU University Amsterdam, De Boelelaan 1081, 1081 HV Amsterdam, The Netherlands*
[2]*Elf Software, Mullerkade 667, 3024 EP Rotterdam, The Netherlands*
[3]*Paderborn University, Theoretical Electrical Engineering, Warburger Straße 100, 33098 Paderborn, Germany*
[4] *TU Dortmund University, Control and Cyber-physical Systems Group, Leonhard-Euler-Str. 2, 44227 Dortmund, Germany*
*\*b.i.avci@vu.nl*

**Abstract:** Here we demonstrate a new type of 3-dB coupler that has an ultra-broadband operational range starting from 1300 nm to 1600 nm with low fabrication sensitivity. The overall device size is 800 µm including in/out S-bend waveguides. The coupler is an asymmetric non-uniform directional coupler that consists of two tapered waveguides. One of the coupler arms is shifted by 100 µm in the propagation direction, which results in a more wavelength-insensitive 3dB-response compared to a standard (not-shifted) coupler. Moreover, compared to a long adiabatic coupler, we achieved a similar wavelength response at a 16-times smaller device length. The couplers were fabricated using the silicon nitride platform of Lionix International. We also experimentally demonstrated an optical switch that is made by using two of these couplers in a Mach Zehnder interferometer configuration. According to experimental results, this optical switch exhibits -10 dB of extinction ratio over a 100 nm wavelength range (1500-1600 nm). Our results indicate that the asymmetric non-uniform directional coupler holds great promise for various applications including optical imaging, telecommunications, and reconfigurable photonic processors where compact, fabrication-tolerant and wavelength-insensitive couplers are essential.

## 1. Introduction

Optical couplers are indispensable components of photonic integrated circuits (PICs), such as multiplexers [1], switches [2], polarization splitters [3], and wavelength filters [4]. Although directional couplers are the commonly used optical splitters due to their structural simplicity, ease of design and zero intrinsic excess loss, their highly wavelength-dependent operation makes them incompatible for use in many PIC applications. Many different approaches have been investigated in order to realize wavelength-insensitive couplers; however, each concept has its own drawbacks [5-15]. One common example is a multi-mode interference coupler [5,6] that couples the input modes to many modes inside the coupler and thereby decreases the wavelength sensitivity by averaging out the uncertainties among these modes. However, typically they exhibit high excess loss. Another example is an asymmetric coupler [14], which is a directional coupler having two coupled waveguides of different widths that exhibits much lower wavelength sensitivity in contrast to conventional symmetric directional couplers. The main drawback of this type of coupler is its high sensitivity to waveguide width variations. However, using curved waveguides, the fabrication sensitivity has been improved significantly [15]. Another form of an asymmetric coupler is an adiabatic coupler [7-10], which exhibits a very wide bandwidth (~400 nm); however, it is rather long (i.e. several millimeters), which prevents its widespread use. Multi-section couplers include two or more directional couplers connected with waveguides having a small path-length difference that can be chosen such that the wavelength dependencies of the directional couplers cancel each other out [11-13]. These couplers are also long and therefore they are not preferable. Considering the drawbacks of existing optical couplers, a compact, ultra-broadband, and fabrication tolerant 3-dB optical coupler is highly needed that can be used both as a stand-alone device and also as the core building block in various applications including optical imaging [16], telecommunications [17], and reconfigurable optical chips and processors [18,19].

In this work, we experimentally demonstrate a new type of a 3-dB optical coupler that works over the 1300-1600 nm wavelength range. The coupler can be considered as an asymmetric non-uniform directional coupler that is comprised of two tapered waveguides with one of them shifted by a distance $L_{shift}$ in the predominant propagation direction (Fig. 1b). As we shall see, this shift reduces the wavelength sensitivity of the coupler at a compact device footprint. The optimum shift value was extracted from a series of beam propagation simulations. To experimentally prove the effect of $L_{shift}$, two couplers were fabricated; one with $L_{shift} = 0$ (not shifted coupler) and one with $L_{shift} = 100$ µm (shifted coupler) using the silicon nitride ($Si_3N_4$) platform of Lionix International [20]. A substantially more wavelength-flattened response was obtained with the shifted coupler. Moreover, we realized an adiabatic coupler using 8-mm-long tapered waveguides. The shifted coupler exhibited a wavelength response similar to the adiabatic coupler at a 16-times-smaller device length. Finally, we fabricated a Mach-Zehnder interferometer (MZI) type optical switch by cascading two of these couplers with a phase section in between. According to experimental results, this optical switch exhibits -10 dB of extinction ratio over the 100 nm wavelength range (1500-1600 nm).

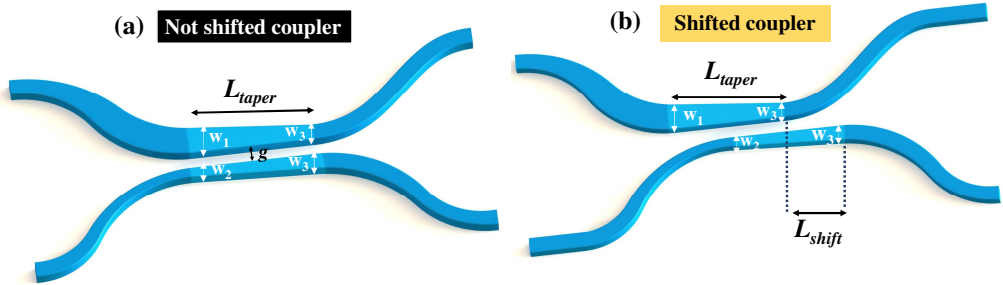

**Fig.1.** Schematic of the coupler **(a)** without shift and **(b)** with shift. Design parameters are indicated in each schematic.

## 2. Methodology

### 3.1 Waveguide geometry and fabrication

All couplers and the optical switch were realized using single-mode $Si_3N_4$ channel waveguides (Lionix International) as given in Fig. 2a. A thin layer of LPCVD-deposited $Si_3N_4$ with a thickness of $t = 185$ nm, and a 4-µm-thick PECVD-deposited silicon dioxide ($SiO_2$) cladding layer (refractive index 1.44) were used to form the channel waveguides. The LPCVD $Si_3N_4$ film was deposited on an 8-µm-thick thermally-oxidized silicon wafer. The refractive index of the thermal oxide and $Si_3N_4$ layer are 1.464 and 2.0 at $\lambda=1550$ nm, respectively. Single-mode optical waveguides with a target width of 1 µm were designed. The effective refractive index of the waveguide at $\lambda=1550$ nm was calculated to be 1.51 using Beam Propagation Method (BPM) simulations (RSOFT Inc.). The mode profile is given in Fig. 2b. We restrict the theoretical discussion and the experiments to single-mode (TE) operation. Material dispersion is included in all simulations, with refractive index data provided by Lionix International. Edge couplers with a waveguide width of 200 nm were used at the input and outputs of the devices in order to reduce the in/out coupling losses. Red light propagation in the straight waveguide is shown in Fig. 2c. For the optical switch, 1-mm-long co-planar gold electrical heaters were used.

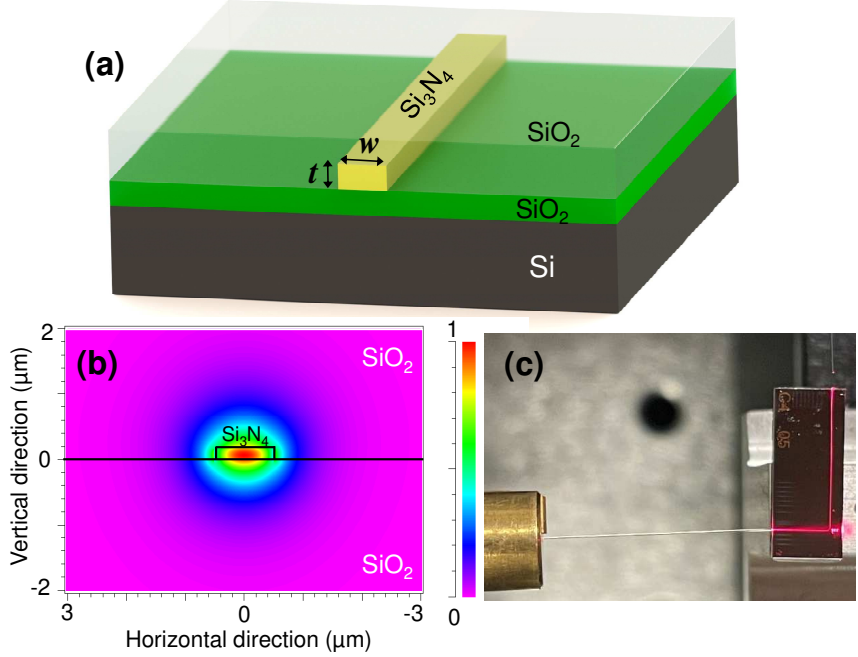

**Fig.2. (a)** Schematic of the optical waveguide geometry. **(b)** Mode profile of the waveguide. **(c)** Red light propagation of the waveguides.

*3.2 Coupler design*

The design of our coupler evolved from the concept of an adiabatic coupler [8], which is known to be the best coupler in terms of wavelength insensitivity with the device size as a major drawback. It consists of two very-long tapered waveguides as depicted in Fig. 3a. If the coupler is sufficiently asynchronous at its input ($X = \Delta\beta/2\kappa > 1$, where X is the asynchronicity parameter and $\Delta\beta$ and $\kappa$ is the difference in the propagation constants of, and the coupling coefficient between, the isolated waveguides, respectively [9]), the power in each of the input waveguides will predominantly go into only one of the system modes. In the coupler discussed here, the synchronicity of the waveguides is arranged in such a way that only the zero-order local system mode of the coupler will be excited. If the widths of the coupled waveguides change adiabatically, i.e. the cross-section of the system changes slowly with the z-position, the particular system mode will propagate through the structure with negligible power conversion to the other system mode. If the widths of the waveguides are interchanged at the end of the tapered coupling region, this gives a full coupler, whereas equal output widths will result in a 3-dB coupler [7]. The theoretical analysis of adiabatic couplers is given in Ref. [8,9]. We first designed an adiabatic coupler with sufficiently long tapered waveguides with $L_{taper}$ = 8 mm. Some design parameters were preset considering the fabrication limitations (gap, g), single mode operation, and compactness criteria. The widths of the input and output waveguides of the tapered sections were chosen as $w_1$ = 1.3 µm, $w_2$ = 1µm, and $w_3 = (w_1 + w_2)/2$ = 1.15 µm, to satisfy single mode operation while minimizing the mode width to reduce fabrication sensitivity. The separation between waveguides was chosen as $g$ = 0.9 µm to avoid possible fabrication limitations while achieving a large enough κ for a smaller footprint. The simulation results given in Fig. 3b (purple line) show the wavelength-flattened response over the 1300-1600 nm wavelength range.

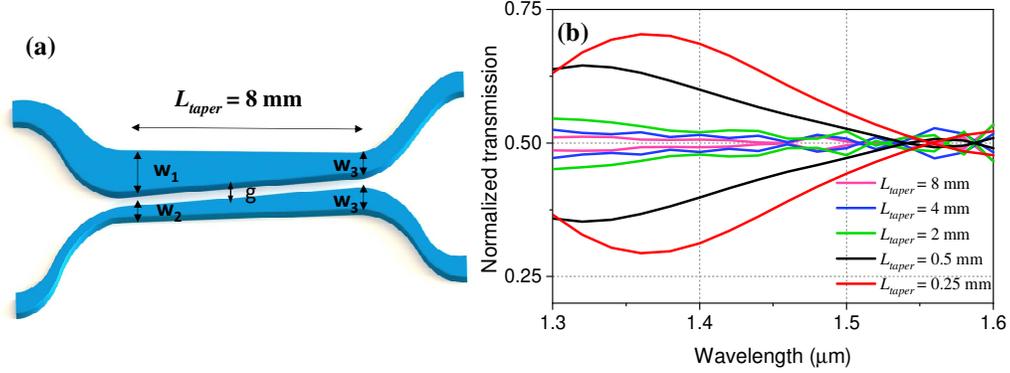

**Fig.3. (a)** The schematic of the long adiabatic coupler ($L_{taper}$ = 8 mm) with relevant parameters. **(b)** The BPM simulation results for different $L_{taper}$ values.

Next, we reduced the length of the tapered region to $L_{taper}$ = 4 mm, 2 mm, 0.5 mm, and 0.25 mm to see the evolution of the wavelength response of the coupler at shorter taper lengths. The results given in Fig. 3b indicate that the wavelength flatness gets worse when the taper length decreases, and at $L_{taper}$ < 0.5 mm the wavelength dependence increases significantly. For $L_{taper}$ > 0.5 mm the coupler still works like a semi-adiabatic coupler with reduced wavelength flatness. In order to have a compact device, we chose the value of $L_{taper}$ = 0.5 mm, which is short enough and not completely non-adiabatic such that there might be a chance that a minor modification leads to a more wavelength-insensitive response. To that end, we implemented the novel approach as depicted in Fig. 1b. The new coupler consists of two tapered waveguides with $L_{taper}$ = 0.5 mm, one of which is shifted in the propagation direction by a certain amount, $L_{shift}$. The optimum shift value was extracted as $L_{shift}$ = 100 µm from the BPM simulation results (Fig. 4a). To do so, the transmission response of the non-uniform coupler was simulated in the range of $L_{shift}$ = 0-150 µm with 5-µm steps. The absolute deviation of the splitting ratio from the 3-dB value was extracted for each $L_{shift}$ value. For the convenience of the reader, only the results corresponding to six $L_{shift}$ values are depicted in Fig. 4a. The transmission responses of the shifted coupler for $L_{shift}$ = 100 µm and the not shifted coupler are given in Fig. 4b. A substantially more wavelength-flattened response was achieved for the shifted coupler with the optimum $L_{shift}$ value.

The shift of the lower branch changes the coupler geometry in several respects: The active region, i.e. the region where the channels are the closest, is shortened. Next, the initial width of the lower waveguide in this active region increases. This implies that the propagation constants of the local modes involved in the interaction — both the modes supported by the individual channels, as well as the supermodes of the composite system — come closer together. Further, there is a slight increase in the gap size along the active region. The last two aspects both influence the strength of the local interaction, partly counteracting each other, such that the net effect of the changes, the effect on the power distribution at the end of the strongest interaction, and in particular its wavelength dependence, is far from obvious. Our coupler can be viewed as a device in-between a long asymmetric adiabatic coupler, and a standard symmetric directional coupler, where neither viewpoint is fully adequate, while the shift of the lower branch alters the mode of action in both descriptions. We thus resorted to the fully numerical design procedure discussed in the former paragraphs.

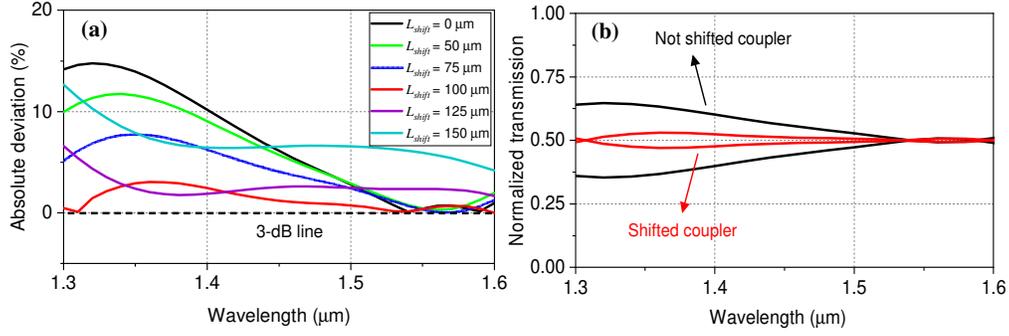

**Fig. 4.** (a) Absolute deviation from 3-dB coupling ratio for different $L_{shift}$ values. Note that for $L_{shift} = 100$ µm, the deviation gets smaller. (b) Simulation results of the shifted and not shifted couplers over the 1300-1600 nm wavelength range.

The evolution of the normalized intensity $I = |E|^2$, where $E$ is the frequency-domain optical electrical field, was calculated using the BPM simulations for the shifted and not shifted couplers. Figure 5 depicts the intensity distribution $I(x,z)$ when the coupler is excited $y = 0$ µm. Apparently, due to $L_{shift}$, the coupling point and the coupling strength change along the propagation direction, which results in a more flattened, close-to-equal splitting ratio for the shifted coupler.

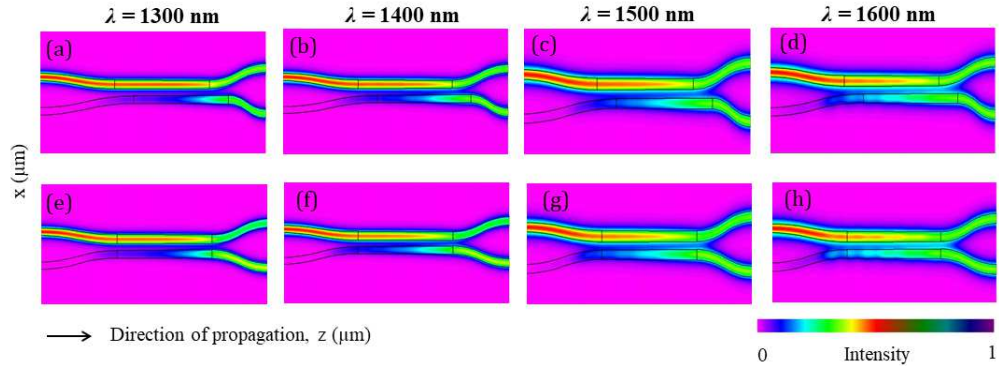

**Fig. 5.** The normalized intensity distribution $I(x,z, y=0)$ of the shifted (a,b,c,d) and not shifted coupler (e,f,g,h) at $\lambda = 1300$ nm, 1400 nm, 1500 nm and 1600 nm.

### 3.3 Parameter tolerance analysis

It is important to check the wavelength-flattened coupler design for its tolerance concerning fabrication errors. Therefore, we investigated the influences of design parameters on the wavelength characteristics. The particular channel waveguide geometry exhibits a high tolerance to the etching process. The LPCVD-deposited $Si_3N_4$ material platform also has a very well-controlled refractive index value, therefore no significant fabrication error is expected on this part. The thickness variation of the $Si_3N_4$ layer is $t < 5$ nm. In this sense, we investigated the effect of changes in the waveguide width and of the $Si_3N_4$ thickness using BPM simulations. The effect of the waveguide width, $w$, (typically ±100 nm) on the transmission response is given in Fig. 6a. It can be seen that the coupling ratio varies at most 3% for a width increase of 100 nm in the 1300 nm range, while it is almost unchanged for the rest of the wavelength range. For a width decrease of 100 nm, the coupling ratio varies at most 5%. Fig. 6b shows the effect of $Si_3N_4$ thickness variation by ± 5 nm. It is found that the coupling ratio stays within the 3 dB ± 0.3 dB range for both thickness changes. In all simulations, only one parameter was varied while keeping the others unchanged.

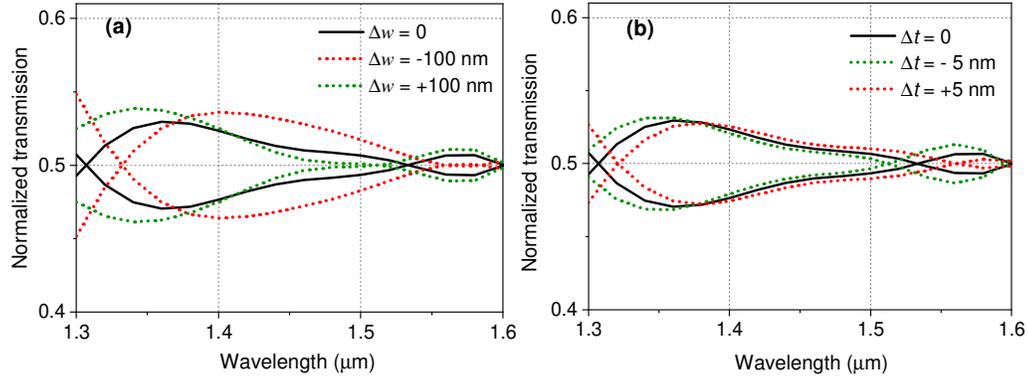

**Fig. 6.** Fabrication tolerance analysis of the non-uniform directional coupler when **(a)** $w$ is varied by ± 100 nm and **(b)** $t$ is varied by ± 5 nm. Note that the wavelength-flattened response of the coupler does not vary more than 5% in both cases.

## 3. Experimental results

In order to compare the improvement provided by $L_{shift}$, two asymmetric non-uniform couplers were designed and fabricated; one with $L_{shift} = 0$ (not shifted coupler, Fig. 1a) and the other one with $L_{shift} = 100$ µm (shifted coupler, Fig. 1b). The overall device size is 800 µm, including in/out S-bend waveguides. Additionally, the adiabatic coupler with $L_{taper}= 8$ mm was fabricated and its response was compared with the shifted coupler's response.

### 3.1 Transmission measurement results of the couplers

Optical transmission measurements were performed by coupling TE-polarized light from a supercontinuum light source (NKT SuperK EXTREME, EXR-4) into the input waveguide with single-mode optical fiber. The output signal was sent to an optical spectrum analyzer (Yokogawa, AQ6370B) through a butt-coupled single-mode fiber. The transmission spectra measured at the output ports of the coupler were normalized with respect to the transmission spectrum of a straight waveguide.

All couplers including the shifted, not shifted and adiabatic couplers were fabricated using the same waveguide platform and their transmission responses were measured using the setup described earlier. The measurement results of the adiabatic coupler are given in Fig. 7a. As expected, a wavelength-flattened response was obtained; however, the taper length was still not sufficient, which can be seen from the oscillatory behavior of the coupler. The transmission responses of the couplers with $L_{shift} = 0$ and $L_{shift} = 100$ µm are given in Fig. 7b and 7c, respectively. According to these results, the optimum shift value of $L_{shift} = 100$ µm improved the wavelength flatness of the coupler significantly without increasing its length. The 3dB±0.3 bandwidth of 300 nm was measured. The excess loss was obtained as 0.5 dB and 1 dB for the shifted and not shifted couplers, by dividing the spectrum of the coupler by the spectrum of a straight waveguide with a similar optical length. Compared with the adiabatic coupler, the shifted coupler exhibits a similar wavelength response at a 16-times-smaller device size.

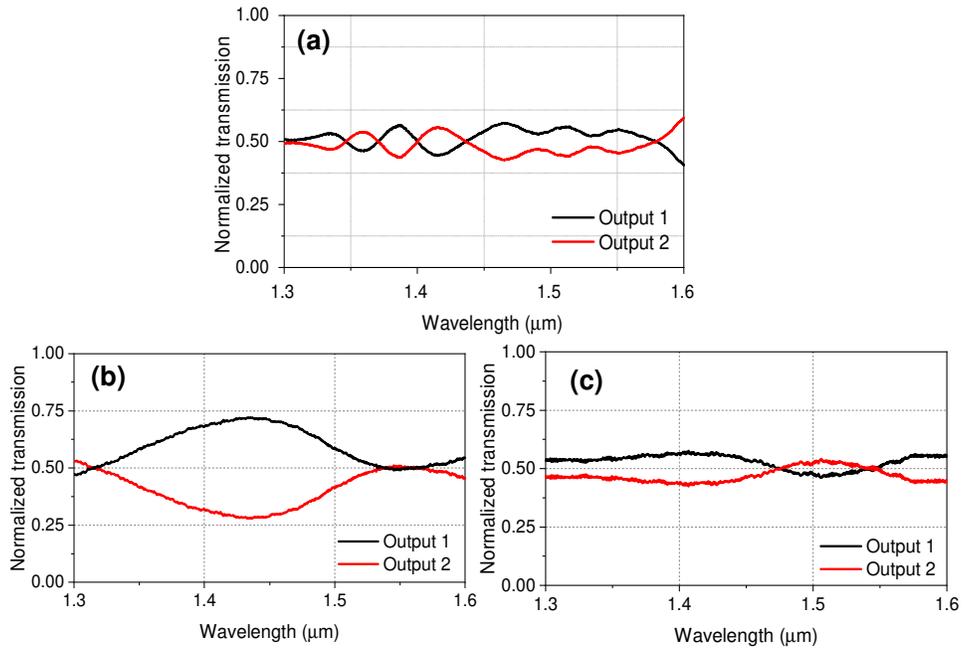

**Fig.7.** The measured transmission response of the **(a)** adiabatic, **(b)** not shifted and **(c)** shifted couplers.

## 3.2 Optical switch results

Broadband optical switches are important components of optical networks and wavelength division multiplexing systems. MZI-type switches are the most preferred version due to their simplicity both in design and fabrication. We formed an MZI-type switch by cascading two non-uniform 3 dB couplers back to back with a phase section (length $L_p$ and phase $\Delta\phi$) on one of the arms as shown in Fig. 8a. When $\Delta\phi = 0$, the light will stay in the bar-port, meaning that the switch is in the *"off"* mode and when $\Delta\phi = \pi$, the light will go through the cross-port, meaning that the switch is in the *"on"* mode. The thermo-optic effect was used to induce the phase difference between the MZI arms. Figure 8b shows the experimental results of the fabricated switch. For a current value of 16.8 mA and a voltage value of 3V, the switch went from the *"off"* to the *"on"* state. We obtained 100 nm bandwidth at a -10dB extinction ratio, which is similar to MZI-type switches reported so far [22-24]. The length of the switch is 2.6 mm including the 0.8 mm length of the couplers and the phase section of $L_p$= 1 mm.

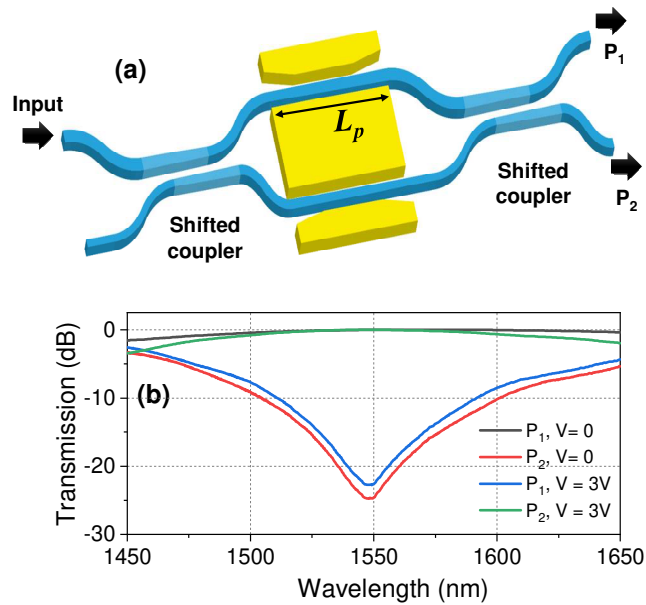

**Fig. 8.** (a) The schematic of the MZI-type switch based on the shifted coupler. $L_p$ is the length of the phase section. (b) The experimental result of the switch. When the applied voltage is V = 3V, the light switches from port $P_1$ to port $P_2$.

## 4. Conclusions

We realized a new type of broadband wavelength-flattened 3-dB optical coupler based on a non-uniform asymmetric directional coupler. A splitting ratio of 3dB±0.3 dB was achieved over the 1300-1600 nm wavelength range. We also compared its performance with a long adiabatic coupler and obtained a similar wavelength response with a 16-times-smaller device length. Using two of these couplers in an MZI configuration, we demonstrated an optical switch with an -10 dB of extinction ratio over the 100 nm wavelength range (1500-1600 nm). We expect that this type of coupler can be successfully applied in many areas of integrated optics, such as optical telecommunication, optical imaging, spectroscopy, programmable photonic chips, or microwave photonics.

**Funding.** ACTPHAST4R (P2019-32), NWO (COMBO-18757) and DFG (231447078–TRR 142, C05).

**Acknowledgments.** The authors thank Ronald Dekker for the fabrication-related discussions.

**Disclosures.** The authors declare no conflicts of interest.

**Data availability.** Data underlying the results presented in this paper are available in Dataset 1, Ref. [25].